\documentstyle[preprint,aps]{revtex}
\tightenlines
\input{psfig}
\begin{document}
\preprint{UNDPDK-98-03e}
\title{$\Delta m^{2}$ Limits from $R(E_{\nu})$}
\author{J.M.~LoSecco}
\address{University of Notre Dame, Notre Dame, Indiana 46556}
\date{\today}
\maketitle
\begin{abstract}
A lower bound on $\Delta m^{2}$ for atmospheric neutrino oscillations
can be obtained from the global value of $R$ at a given energy.
This bound may be more reliable than one based on the observed isotropy.
Some global values of $R$ imply an anisotropy in the neutrino flux that
may, or may not be manifest in the observed neutrino interactions.
A limit of $\Delta m^{2}>0.0077$ eV$^{2}$ is easily obtained.
Much higher mass scales are implied by the reported energy independence
and magnitude of $R$\\

Subject headings: Cosmic Rays --- Elementary Particles --- Neutrino Oscillations
\\
\end{abstract}


The recent announcement of the discovery of neutrino oscillations by the
Super-Kamiokande collaboration\cite{discov} does not, under closer
scrutiny, seem to be consistent with the observations.  The oscillation
length they infer corresponds to 1000's of kilometers at their energies.
It has been pointed out\cite{losecco1} that since the oscillation length
is large compared to half the distance scales in the problem and small
compared with the other half that a serious up-down anisotropy would
be expected in the data.  Such a large anisotropy is not observed.

It has been suggested that the apparent near isotropy may be an
experimental artifact, a consequence of the scattering angle between the
neutrino direction and the reconstructed lepton, compounded by experimental
resolution.  While such angular resolution effects should decrease with energy
(which unfortunately couples the neutrino mass scale to resolution effects)
in this paper we explore what can be learned if the directional information
is ignored.

We employ the variable $R$ commonly defined to be:
\[
R = \frac{(\mu / e )_{DATA}}{(\mu / e )_{MC}}
\]
and often interpreted as the fraction of muon neutrinos observed relative
the expected number.  A value of $R$ less than one is often taken as evidence
for neutrino oscillations and is modeled by:
\[
P_{\nu_{\mu}\rightarrow\nu_{\tau}}=\sin^{2}(2 \theta)\sin^{2}(1.27 \Delta m^{2} \frac{L}{E})
\]
$\sin^{2}(2 \theta)$ and $\Delta m^{2}$ are the two parameters of the model.
$L$ is the distance, in kilometers the neutrino has traveled, $E$ is the
neutrino energy, in GeV and $\Delta m^{2}$ is measured in eV$^{2}$.

Only projections of $R$ as a function of energy and of the cosine of the
zenith angle have been published.  It is possible that the observed isotropy
is a consequence of the large event rate at low energies where the
reconstructed track direction is less well correlated with the initial
neutrino direction.  This paper points out that isotropy of the neutrino flux
within the neutrino oscillation hypothesis is a {\em consequence} of the small
value of $R$ observed and the bimodal nature of the path length scales in
the problem.  Once $R$ gets below about 0.73 the upper hemisphere (downward
going) flux must participate in the oscillations.  Due to the narrow range
of distances for the downward component most of the solid angle will
participate in oscillations.

\begin{figure}
\psfig{figure=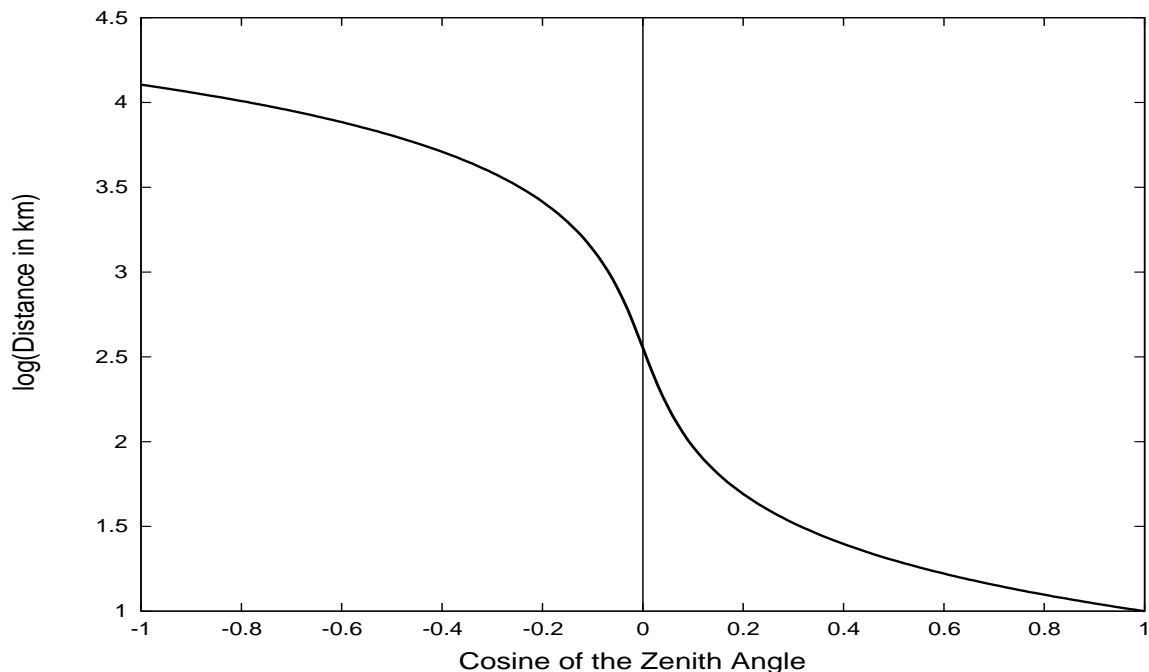,width=6.0in,height=3.50in}
\caption{\label{loglen} Approximate $\log_{10}$ of the neutrino flight path
(in kilometers) as a function of the cosine of the zenith angle.}
\end{figure}

Underground detectors can observe atmospheric neutrinos.  For atmospheric
neutrinos the distance the neutrino has traveled depends on the direction
it enters the detector.  Since the detectors are located near the surface
of the Earth path lengths vary from a few kilometers to about 13,000
kilometers.  Figure \ref{loglen} is a plot of the log of the path length as a
function of neutrino direction for neutrinos originating in the atmosphere.
The path length can be approximated by:
\[
L(\cos(\theta_{z})) = \sqrt{R_{1}^{2} (\cos^{2}(\theta_{z})-1) + R_{2}^{2}}
-R_{1} \cos(\theta_{z})
\]
with $R_{2}$ representing the distance from the
center of the earth to the upper atmosphere where the neutrinos are born and
$R_{1}$ representing the distance from the center of the earth to the detector
($R_{2}>R_{1}$).  Figure \ref{loglen} shows that atmospheric neutrino
oscillations are dominated by two distance scales.  Neutrinos from above have
path lengths of 10's of kilometers.  Those coming from below have path
lengths of the order of 1000's of kilometers.

Neutrinos that have converted from muon type to tau type or a sterile type
would not be observed in the detector.  This reduction of the flux of
muon neutrinos, relative to electron neutrinos would be manifest by a value
of $R<1$.  Low values of $R$ have been observed\cite{all,imb92}.

One can obtain more information about the nature of the oscillations by
measuring $R$ in specific energy and/or directional bins.  But at very low
energies the direction of the reconstructed event may not be closely
associated with the direction of the initial neutrino and so one may be
averaging over substantial distances.  Distance varies slowly with angle near
the vertical so this is not necessarily a serious problem.

The global value of $R$ (for an isotropic atmospheric source) is given by
\[
R=1-\frac{1}{2} \times \int_{-1}^{1} \sin^{2}(2 \theta) \sin^{2}(1.27 \Delta m^{2}
\frac{\sqrt{R_{1}^{2} (\cos^{2}(\theta_{z})-1) + R_{2}^{2}} 
-R_{1} \cos(\theta_{z})}{E}) d \cos(\theta_{z})
\]
Which is a function of $\Delta m^{2}/E$

\begin{figure}
\psfig{figure=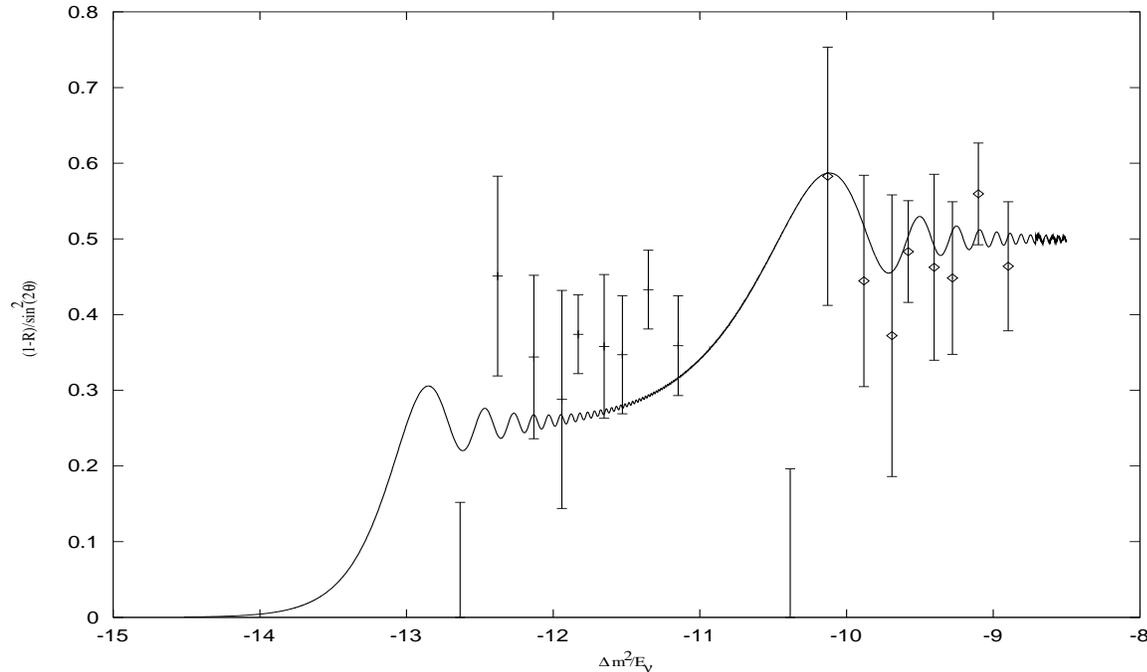,width=6.0in,height=3.50in,angle=-90}
\caption{\label{rlim} $\frac{1-R}{\sin^{2}(2 \theta)}$ plotted as a function
of the $\log_{10}$ of $\Delta m^{2}/E_{\nu}$ with $\Delta m^{2}$ in eV$^{2}$
and E measured in eV.  $\frac{1-R}{\sin^{2}(2 \theta)}$ is
the maximum global fraction of muon neutrinos that
have transformed into a noninteracting species when integrated over all
atmospheric neutrino path lengths.
The points plotted near the center of the figure are from references
5 and 6 using the $\Delta m^{2}$ from reference 2.  The points plotted to
the right are our fit to $\Delta m^{2}$ using the data of references 5 and 6.
}
\end{figure}

Since $\sin^{2}(2 \theta) \le 1$ one can get a bound on 
$\Delta m^{2}/E$ from the observed value of $R$ at a given energy.
Figure \ref{rlim} plots $\frac{1-R}{\sin^{2}(2 \theta)}$ as a function of
$\Delta m^{2}/E$ when the flux is integrated
over all solid angles.  $\frac{1-R}{\sin^{2}(2 \theta)}$ is an upper bound
on $1-R$.  At low values of $\Delta m^{2}/E$ the neutrinos have
not had time to oscillate so $1-R$ is zero.  At very large values of
$\Delta m^{2}/E$ the neutrinos traveling over all paths in the problem have
gone through several oscillations so that $\frac{1-R}{\sin^{2}(2 \theta)}$
is $\frac{1}{2}$.
The interesting structure near the central region of the plot comes about
as the oscillation length becomes comparable to various distances in the
problem.  The plateau at about 0.25 comes about for that range of oscillation
lengths for which the downward path is too short for significant
changes to have occurred but where the upward direction is large compared
to the oscillation length.
A value of $R$ in the intermediate range of $0.5 < R < 0.73$
implies a flux isotropy since the upper hemisphere must participate in
oscillations to get $R$ this small.
One can still get isotropy with a large $R$ for small values of
$\sin^{2}(2 \theta)$.

The observed value of $R$\cite{ratio} for all contained events is
0.61$\pm$0.03$\pm$0.05 ($R < 0.68$ at 90\% confidence limit).
For this work we need the value of R at a {\em specific} energy.  From figure 4
of reference \cite{ratio} it can be found that $R = 0.63 \pm 0.05$ at 1.2 GeV.
This value
gives an upper bound of $R < 0.69$ at 90\% confidence level.
As seen from figure \ref{rlim} for this value of $R$ that
$\log_{10}(\Delta m^{2}/E)$ is greater than about -11.19 in units of eV
or -2.19 in units of eV$^{2}$/GeV.
At 1.2 GeV this gives a bound of
$\Delta m^{2} > 10^{-2.19}$ eV$^{2}$/GeV $\times 1.2$ GeV
$ > 7.7 \times 10^{-3}$ eV$^{2}$.

The observed values of $R$ are energy independent over a substantial range
\cite{ratio,multi}.
From figure \ref{rlim} it can be seen that $R$ is expected to be
independent of energy over a
substantial range.  In particular for $\Delta m^{2}/E < 10^{-13.8}$ eV
$R$ will be energy independent at a value near one.
For $\Delta m^{2}/E > 10^{-9.8}$ eV $R$ will be energy independent with a
minimum value near $\frac{1}{2}$.  The intermediate range plateau
$10^{-13.0} < \Delta m^{2}/E < 10^{-12.0}$ eV would also appear to be
energy independent with suitable energy averaging.  In this intermediate
range the observations would have a noticeable
``neutrino twilight''\cite{losecco1} with the upper hemisphere being
considerably brighter than the lower one. 

The 5 points in reference \cite{ratio} and the 4 multi GeV points from
\cite{multi} may be fit to find the best values of $\sin^{2}(2 \theta)$ and
$\Delta m^{2}$.  Physical values of $\sin^{2}(2 \theta)$ must be bounded
in the range $0 < \sin^{2}(2 \theta) < 1$.  A formal fit is difficult since the
errors in references \cite{ratio} and \cite{multi}
are dominated by systematic effects and are correlated.  So it is
difficult to get a true confidence level associated with the fit.
From the global value of $R$ cited above it can be inferred
that the common systematic error is on the order of 0.047.  Removing this
common factor and doing a $\chi^{2}$ minimization with the resulting
reduced errors yields a best fit to $\sin^{2}(2 \theta) = 0.8$
and $\Delta m^{2}=0.24$ eV$^{2}$.  The results of the fit are the
points plotted on the right of figure \ref{rlim}.  In actual fact there
is no best fit since all values above about $\Delta m^{2} > 0.1$ eV$^{2}$
give a reasonably good fit to the data.

For reference
the points plotted near the center of the figure are the Super Kamioka
``best fit''
to the neutrino oscillation hypothesis, $\sin^{2}(2 \theta)=1$ and
$\Delta m^{2} = 2.2 \times 10^{-3}$.  The Super Kamioka fit could be improved
if the points could be lowered to center on the curve.  But the
$\sin^{2}(2 \theta)=1$ bound has been reached and the points can not be
lowered.  To get good agreement for $\Delta m^{2} = 2.2 \times 10^{-3}$
would require a value of $\sin^{2}(2 \theta)=1.5$ which is unphysical.

(Note:  It is the same data from Super Kamioka plotted at the center and the
right of figure \ref{rlim}.  They appear in different places since the
parameters of the fit $\sin^{2}(2 \theta)$ and $\Delta m^{2}$ have been
incorporated into the plot axes.)

It is difficult to quantify the confidence level associated with this fit
of $\Delta m^{2}$ to $R(E)$ for all 9 points.
In an IMB presentation of the atmospheric neutrino
anomaly\cite{imb92} the common systematic errors are clearly assigned to
the expected values and the measured values are illustrated with the truly
independent statistical errors.

Our relation above for $R(\frac{\Delta m^{2}}{E_{\nu}})$ assumed that the
source was isotropic.  That is, it was assumed that the neutrinos were
uniformly distributed over the path length distribution of figure \ref{loglen}.
At modest energies the neutrino flux may be modulated by geomagnetic effects
that restrict the number of primary cosmic ray particles that can impact the
atmosphere and create neutrinos.  This would modulate the path length
distribution.  In the $U-D$ analysis of
reference \cite{discov} (which is substantially the same data as in
\cite{ratio,multi}) it is indicated that such geomagnetic effects are less than
2\% at sub-GeV energies and less than 1\% at higher energies.

To get agreement with the observed bound of $R<0.68$
with a $\Delta m^{2}$ of the order of $2 \times 10^{-3}$ eV$^{2}$
would require a large neutrino flux anisotropy.
An expected upward going muon neutrino flux approximately 80\% greater than
the downward flux (1.8 times) is needed.
While neutrino oscillations would remove much (50\%) of the excess upward muon
neutrino flux the electron neutrino signal would
reflect this serious distortion.  To achieve a value of $R=0.61$ would require
3.5 times greater upward neutrino flux.

In the absence of very large directional modulation in the parent neutrino flux
the upper hemisphere needs to participate in neutrino oscillations to achieve
such a small value of $R$ via this simple two component oscillation mechanism.

\section*{Acknowledgments}
I would like to thank F.~Vissani for some helpful questions about the
anisotropy.  I would like to thank Bill Shephard and John Poirier for
careful reading of the manuscript.  Jim Wilson has helped confirm some of
the numerical results and provided much encouragement.

\begin{references}
\bibitem{discov} Y.~Fukuda {\em et al.},  Phys.~Rev.~Lett. {\bf 81}, 1562
(1998).
T.~Kajita ``Atmospheric Neutrino Results from Super-Kamiokande and
Kamiokande -- Evidence for $\nu_{\mu}$ Oscillations'', Talk presented at
Neutrino '98, Takayama Japan, June 4, 1998.
\bibitem{losecco1} J.M.~LoSecco ``Problems with Atmospheric Neutrino
Oscillations'',  hep-ph/9807359 (to be published).
\bibitem{all} T.J.~Haines {\em et al.}, Phys.~Rev.~Lett. {\bf 57}, 1986
(1986).\\
K.S.~Hirata {\em et al.}, Phys.~Lett. {\bf B205}, 416 (1988).\\
D.~Casper {\em et al.}, Phys.~Rev.~Lett. {\bf 66}, 2561 (1991).
\bibitem{imb92} R.~Becker-Szendy {\em et al.}, Phys.~Rev. {\bf D46}, 3720
(1992).
\bibitem{ratio} Y.~Fukuda {\em et al.},  
Phys.~Lett. {\bf B433}, 9 (1998).
\bibitem{multi} Y.~Fukuda {\em et al.},
Phys.~Lett. {\bf B436}, 3 (1998).
\end{references}
\end{document}